\documentclass[aps,prd,longbibliography,nofootinbib,amsmath,amssymb,amsfonts,notitlepage,superscriptaddress]{revtex4-1}
\usepackage[utf8]{inputenc}
\usepackage[T1]{fontenc} 
\usepackage[english]{babel}
\usepackage{graphicx} %imagens
\usepackage{float}    %imagens
\usepackage{hyperref}
\usepackage{cleveref} %refs
\usepackage{amsfonts,amsthm}
\usepackage{amsmath,amssymb}
\usepackage[usenames,dvipsnames]{xcolor}

\newcommand{\be}{\begin{equation}}
\newcommand{\ee}{\end{equation}}
\newcommand{\bea}{\begin{eqnarray}}
\newcommand{\eea}{\end{eqnarray}}

\newcommand{\Z}{\mathbb{Z}}
\date{\today}

\begin{document}
\date{\today}
\title{
{\normalsize \hfill CFTP/19-022} \\*[7mm]
Boundedness from below in the $U(1)\times U(1)$ Three-Higgs-Doublet model}
\author{Francisco S. Faro}
\email{francisco.faro@tecnico.ulisboa.pt}
\author{Igor P. Ivanov}
\email{igor.ivanov@tecnico.ulisboa.pt}
\affiliation{Centro de Física Teórica de Particulas, Departamento de Física, Instituto Superior Técnico, Universidade de Lisboa, Lisboa}

\begin{abstract}
Establishing if multi-Higgs potentials are bounded from below (BFB) can be rather challenging, and it may impede efficient investigation of all phenomenological consequences of such models. In this paper, we find the necessary and sufficient BFB conditions for the Three-Higgs-Doublet model (3HDM) with the global symmetry group $U(1)\times U(1)$. We observed an important role played by charge-breaking directions in the Higgs space, even for situations when a good-looking neutral minimum exists.
This remark is not limited to the particular model we consider but represents a rather general feature of elaborate multi-Higgs potentials which must be carefully dealt with.  
Also, applying this method to Weinberg's model (the $\Z_2 \times \Z_2$ symmetric 3HDM) turned out to be more challenging than was believed in the literature. In particular, we have found that the approach taken in a paper from 2009 does not lead to the necessary and sufficient BFB conditions for this case.
\end{abstract}

\maketitle

\section{Introduction}

Experimental data do not force the Higgs sector to be as minimal as postulated in the Standard Model (SM). Building models with non-minimal Higgs sectors is an attractive option \cite{ivanov2017building}, as it helps to resolve various open problems, both in particle physics and in cosmology, which the SM cannot address. When building these models, one has to deal with technical issues related to the properties of the scalar potential; in particular, one has to ascertain that it is bounded from below (BFB) so that a vacuum state exists at all. 

Even for sophisticated scalar sectors, it is often easy to give a set of {\em sufficient} BFB conditions. Models satisfying them are safe and can be used in phenomenological analyses. However, such conditions can be overly restrictive, leaving out parts of the parameter space with potentially intriguing phenomenological consequences. Thus, when exploring the parameter space in a class of multi-Higgs models, it is always desirable to establish the exact BFB conditions which are simultaneously necessary and sufficient. This technical issue is rather challenging and has only been solved for sufficiently simple cases. For example, in models with $N$ Higgs doublets (NHDMs), the exact BFB conditions are known for the general 2HDM \cite{Maniatis:2006fs,Ivanov:2006yq,Nishi:2007nh} and for several versions of NHDMs equipped with various global symmetries \cite{klimenko1985conditions,grzadkowski2009natural,kannike2012vacuum,Degee:2012sk,kannike2016vacuum}. In other cases, general strategies were outlined \cite{Maniatis:2015gma,ivanov2018algorithmic} but they did not yet result in a closed set of BFB conditions in terms of the parameters of the potential. In fact, it is well possible that, starting from sufficiently sophisticated cases, it may be impossible to present these conditions in the form of algebraic inequalities \cite{ivanov2018algorithmic}.

In this paper, we report the necessary and sufficient BFB conditions for yet another model, the 3HDM with global symmetry group $U(1)\times U(1)$. This model is, in fact, a particular case of the original Weinberg's model with symmetry group $\Z_2 \times \Z_2$ which sparked the 3HDM activity back in 1976 \cite{weinberg1976gauge}. It is instructive to mention that at that time the main focus was on phenomenological aspects of the model; the stability of the potential was {\em assumed}, with little attention paid to the exact BFB conditions \cite{weinberg1976gauge,branco1980spontaneous}. It was only in 2009 that a set of the BFB conditions for the original $\Z_2 \times \Z_2$ was published \cite{grzadkowski2009natural}. Since the $U(1)\times U(1)$ 3HDM considered here is a particular case of Weinberg's model, one would expect that its BFB conditions should emerge from the previous results in a limiting case. We found our results not to meet this expectation, and we will discuss the origin of this discrepancy. When deriving the conditions for the $U(1)\times U(1)$-symmetric 3HDM, we learned yet another lesson: one must always check stability along charge-breaking directions in the Higgs space, even if one has a normally looking neutral minimum. We will show an example in which such a minimum exists and the potential is stable in all {\em neutral} directions, but it is unbounded from below along some charge-breaking directions.

The paper is organised as follows. In section 2 we describe the $U(1) \times U(1)$ symmetric 3HDM and find the necessary and sufficient BFB conditions. We first remind the reader of the copositivity conditions and then adapt the potential to the form in which these conditions are applicable. When deriving BFB conditions, we pay special attention to the charge-breaking directions in the Higgs space. In section 3 we further investigate these results and draw from them two lessons, which we find important to communicate. First, we highlight the importance of charge breaking directions. Second, turning to the case of the $\Z_2 \times \Z_2$ symmetric 3HDM, we find it more challenging than previously described in the literature. Thus, establishing the exact BFB conditions in this case is still an open question.

\section{Necessary and sufficient BFB conditions for the $U(1)\times U(1)$ 3HDM}
\subsection{Copositivity conditions}\label{subsection-copositivity}

We begin by reminding the reader of the structure of the Higgs space in the 3HDM and of the {\em copositivity conditions} \cite{kannike2012vacuum}, a simple but powerful approach to establish the BFB conditions when the scalar potential can be written as a quadratic form of positive definite variables. The Higgs space of the 3HDM is spanned by three $SU(2)_L\times U(1)_Y$ scalar doublets with hypercharge $Y=1/2$, and we make use of the standard definition for the electric charge, where the upper components of the doublets are charged. When analysing the structure of the potential, we replace the scalar field operators $\phi_i(x)$ by $c$-numbers which can be conveniently parameterized as
\begin{equation}
    \phi_i=\sqrt{r_i}e^{i\gamma_i}\begin{pmatrix} \sin{\left(\alpha_i\right)} \\ \cos{\left(\alpha_i\right)}e^{i\beta_i} \end{pmatrix}, \quad i=1,\,2,\,3.
\end{equation}
Here the norms of the doublets are represented by $(\phi^\dagger_i\phi_i)= r_i \ge 0$, which we shall refer to as the radial variables. The angles $\alpha_i$ and the phases $\beta_i$, $\gamma_i$ are called angular variables. By allowing for any values of the phases, we can restrict $\alpha_i$ to lie within the first quadrant without losing generality. {\em Neutral directions} in the Higgs space correspond to situations when all $\phi_i$ are proportional to each other. One can define the non-negative charge-breaking sensitive quantities
\begin{equation}
\begin{split}
    z_{ij}=&\ (\phi^\dagger_i\phi_i)(\phi^\dagger_j\phi_j) - (\phi^\dagger_i\phi_j)(\phi^\dagger_j\phi_i) \\
    =&\  r_i r_j \left[\sin^2\!\alpha_i\cos^2\!\alpha_j + \sin^2\!\alpha_j\cos^2\!\alpha_i - 2\sin\alpha_i\cos\alpha_i \sin\alpha_j\cos\alpha_j\cos(\beta_i - \beta_j) \right] \ge 0\,, \label{zij}
\end{split}
\end{equation}
and notice that taking all $z_{ij}=0$ corresponds to neutral directions of the Higgs space. Other directions, along which the strict proportionality of all three doublets does not hold, are called {\em charge breaking} directions and they correspond to at least one $z_{ij} \neq 0$. One can clearly see this by applying the $SU(2)\times U(1)$ gauge transformation to all three doublets, rewriting them as
\begin{equation} \label{eq:Igor_parameterization}
\begin{split}
    \phi_1=\sqrt{r_1}\begin{pmatrix} 0 \\ 1 \end{pmatrix},\quad \phi_2=\sqrt{r_2}\begin{pmatrix} \sin{\left(\alpha_2\right)} \\ \cos{\left(\alpha_2\right)}e^{i\beta_2} \end{pmatrix}, \quad \phi_3=\sqrt{r_3}e^{i\gamma}\begin{pmatrix} \sin{\left(\alpha_3\right)} \\ \cos{\left(\alpha_3\right)}e^{i\beta_3} \end{pmatrix}\,,
\end{split}
\end{equation}
so that $\alpha_{2,\,3}$ can be identified as the charge breaking angles.

The global group $U(1)\times U(1)$ can be represented, in a suitable basis, by arbitrary rephasing transformations of individual doublets. The 3HDM potential invariant under them can be written as $V = V_2 + V_N + V_{CB}$, where
\begin{align}
V_2 &=  m_{11}^2(\phi_1^\dagger\phi_1) + m_{22}^2(\phi_2^\dagger\phi_2) + m_{33}^2(\phi_3^\dagger\phi_3)\,, 
\label{V2}\\
V_N &= \frac{\lambda_{11}}{2}(\phi_1^\dagger\phi_1)^2 + \frac{\lambda_{22}}{2}(\phi_2^\dagger\phi_2)^2 + \frac{\lambda_{33}}{2}(\phi_3^\dagger\phi_3)^2 + \lambda_{12}(\phi_1^\dagger\phi_1)(\phi_2^\dagger\phi_2) + \lambda_{13}(\phi_1^\dagger\phi_1)(\phi_3^\dagger\phi_3) + \lambda_{23}(\phi_2^\dagger\phi_2)(\phi_3^\dagger\phi_3) \,,\label{VN1}\\
V_{CB} &= \lambda'_{12} z_{12} + \lambda'_{13} z_{13} + \lambda'_{23}z_{23}\,. \label{VCB1}
\end{align}
Along neutral directions all $z_{ij}=0$, so that the scalar potential is given by $V= V_2 + V_N$. Along charge breaking directions, at least one $z_{ij}\neq0$, meaning that $V_{CB}$ switches on and contributes to the potential. 

Let us now focus on the potential along neutral directions and establish conditions for its boundedness from below. As usual, this is equivalent to requiring that the quartic part of the potential is non-negative along all neutral directions. Using the parameterization in \eqref{eq:Igor_parameterization}, we express the potential as a quadratic form of the variables $r_i \ge 0$:
\bea
    V_N &=&  \frac{\lambda_{11}}{2}r^2_1 + \frac{\lambda_{22}}{2}r^2_2 + \frac{\lambda_{33}}{2}r^2_3 + \lambda_{12}r_1r_2 + \lambda_{13}r_1r_3 + \lambda_{23}r_2r_3 \label{VN2} \\
    &\equiv & \frac{1}{2} A_{ij}r_i r_j\,.\label{generic-A}
\eea
In order for the $3\times 3$ real symmetric matrix $A$ to be positive definite (or, at least, non-negative) in the first octant of variables $r_i$, its entries must satisfy the following list of inequalities known as the {\em copositivity conditions} \cite{kannike2012vacuum}:
\begin{equation}
    \begin{split}
        &A_{11} \geq 0,\quad A_{22} \geq 0,\quad A_{33} \geq 0, \\
        &\bar{A}_{12} \equiv \sqrt{A_{11}A_{22}} + A_{12} \geq 0,\\
        &\bar{A}_{13} \equiv \sqrt{A_{11}A_{33}} + A_{13} \geq 0,\\
        &\bar{A}_{23} \equiv \sqrt{A_{22}A_{33}} + A_{23} \geq 0,\label{copositivity1}
    \end{split}
\end{equation}
 and
\begin{gather}
    \sqrt{A_{11}A_{22}A_{33}} + A_{12}\sqrt{A_{33}} + A_{13}\sqrt{A_{22}} + A_{23}\sqrt{A_{11}} + \sqrt{2\bar{A}_{12}\bar{A}_{13}\bar{A}_{23}} \geq 0\,. \label{copositivity2}
\end{gather}
These are the necessary and sufficient conditions for $V_N$ to be bounded from below.

\subsection{Including charge-breaking directions}
The charge breaking part of the potential depends not only on the radial variables $r_i$, but also on the angular variables $\alpha_i,\,\beta_i,\,\gamma_i$. Thus, the quartic part of the potential, $V_4$, cannot yet be written as a quadratic form of independent and non-negative variables.\footnote{Although $z_{ij} \ge 0$, they are {\em not} independent from $r_i$.} However, we can find the directions along which $V_{CB}$ reaches a minimum in the angular variables. If $V_{CB}$ evaluated along these special directions can be written as a quadratic form of $r_i$, then the quartic potential $V_N + V_{CB}$ can still be written in the same generic form as in eq.~\eqref{generic-A}, where the matrix $A$ receives, in addition to \eqref{VN2}, contributions from $V_{CB}$. Then, one just applies the same copositivity conditions in \eqref{copositivity1} and \eqref{copositivity2} to the total matrix $A$. The resulting conditions can be stronger than just for $V_N$, but only in the case when $V_{CB}$ can become negative along some angular directions.

By differentiating \cref{VCB1} with respect to $\delta \equiv \beta_3 - \beta_2$
\begin{equation}
    2\frac{\partial V_{CB}}{\partial\delta}=\lambda_{23}^{\prime}r_2r_3\sin{\left(2\alpha_2\right)}\sin{\left(2\alpha_3\right)}\sin{\left(\delta\right)}=0,    
\end{equation}
we obtain the extrema at $\delta=0$ and $\pi$. At these values of $\delta$, the potential is
\begin{equation}\label{eq:V_CB}
    V_{CB}= \lambda_{12}^{\prime}r_1r_2\sin^2{\left(\alpha_2\right)} + \lambda_{13}^{\prime}r_1r_3\sin^2{\left(\alpha_3\right)} + \lambda_{23}^{\prime}r_2r_3\sin^2{\left(\alpha_2 \mp \alpha_3\right)}.
\end{equation} 
Next, we find at which values of angles $\alpha_2$ and $\alpha_3$, but with fixed $r_i$, this expression takes its minimal value. Clearly, we are interested in cases when $V_{CB}$ can reach negative values, helping to lower the overall quartic potential, which is possible whenever there is at least one negative $\lambda'_{ij}$. To treat all cases in a uniform fashion, we write $\lambda^\prime_{ij}=\sigma_{ij}|\lambda^\prime_{ij}|$ and keep track of the sign factors $\sigma_{ij}=\pm1$. Moreover, notice that if one $r_i$ is zero, for example $r_1=0$, the problem is solved immediately: the minimum value of $V_{CB}$ is either zero for $\sigma_{23} = +1$ or $-|\lambda_{23}^{\prime}|r_2r_3$ for $\sigma_{23} = -1$. Thus, we consider below only the case when all $r_i \not = 0$.

Here we shall sample the whole range of the angles $\alpha_2$ and $\alpha_3$, then the sign choice in $\alpha_2 \mp \alpha_3$ does not matter. Setting the derivatives of eq.~\eqref{eq:V_CB} with respect to $\alpha_2$ and $\alpha_3$ to zero, we arrive at two simultaneous equalities
\begin{gather}
    \frac{|\lambda^\prime_{23}|}{r_1}\sin{\left[2\sigma_{23}\left(\alpha_2 - \alpha_3\right)\right]}=\frac{|\lambda^{\prime}_{13}|}{r_2}\sin{\left(2\sigma_{13}\alpha_3\right)} = \frac{|\lambda^{\prime}_{12}|}{r_3}\sin{\left(-2\sigma_{12}\alpha_2\right)}\,.
    \label{two-equalities}
\end{gather}
These two equations can have trivial and non-trivial solutions. Trivial solutions arise when we require that the quantity in \eqref{two-equalities} is equal to zero. They correspond to $\alpha_2$ and $\alpha_3$ equal to multiplies of $\pi/2$. Substituting these values in the potential of eq.~\eqref{eq:V_CB}, one obtains three angular extrema of $V_{CB}$:
\begin{align}
    \lambda_{12}^{\prime}r_1r_2 + \lambda_{23}^{\prime}r_2r_3\,,\quad 
    \lambda_{13}^{\prime}r_1r_3 + \lambda_{23}^{\prime}r_2r_3\,, \quad 
    \lambda_{12}^{\prime}r_1r_2 + \lambda_{13}^{\prime}r_1r_3\,.\label{VCB-triv}
\end{align}
Clearly, it only makes sense to include those expressions for situations in which at least one of $\lambda'_{ij} < 0$.

When looking for non-trivial solutions, we observe that equations \eqref{two-equalities} resemble the law of sines in a triangle\footnote{This triangle technique was already used by Branco \cite{branco1980spontaneous,branco1999cp} to find $CP$ breaking minima for the real $\Z_2 \times \Z_2$ symmetric 3HDM, the so-called Branco's model.}:
\be
\frac{\sin\theta_1}{L_1} = \frac{\sin\theta_2}{L_2}  = \frac{\sin\theta_3}{L_3} \,.\label{generic-triangle}
\ee
In order for this analogy to work, the lengths of the triangle
\begin{gather}
    L_1=\frac{r_1}{|\lambda^{\prime}_{23}|}, \quad L_2=\frac{r_2}{|\lambda^{\prime}_{13}|}, \quad L_3=\frac{r_3}{|\lambda^{\prime}_{12}|}\label{Li}
\end{gather}
must satisfy the well-known triangle inequalities, 
%\footnote{The sum of the lengths of two sides must be bigger than, or equal to, the length of the remaining side.},
\begin{gather}
    \left|L_1 - L_2\right| \le L_3 \le L_1 + L_2\,, \label{eq:prism}
\end{gather}
and the inner angles of the triangle must sum up to $\theta_1 + \theta_2 + \theta_3 = \pi$. The exact relations between $\theta$'s and $\alpha$'s depend on sign factors $\sigma_{ij}$. For example, for $\sigma_{12}=\sigma_{13}=\sigma_{23}=+1$, we set 
\begin{equation}
    \theta_1= \pi-2\left(\alpha_2 - \alpha_3\right), \quad \theta_2=\pi-2\alpha_3, \quad \theta_3=-\pi + 2\alpha_2\,,
\end{equation}
while for $\sigma_{13}=\sigma_{13}=+1$ and $\sigma_{12} = -1$, we set
\begin{equation}
    \theta_1= 2\left(\alpha_2 - \alpha_3\right), \quad \theta_2=2\alpha_3, \quad \theta_3=\pi - 2\alpha_2\,.
\end{equation}
Hence, if the triangle inequalities in \eqref{eq:prism} are satisfied, there always exists a (unique) solution for angles $\alpha_i$. It is remarkable that, when these values are substituted back in \eqref{eq:V_CB}, they produce, once again, a quadratic form in the variables $r_i$. The value of $V_{CB}$ can be expressed in a compact form
\be
    V_{CB}^{\rm non-triv.}=\frac{\lambda^{'}_{12}\lambda^{'}_{23}\lambda^{'}_{31}}{4}
    (L_1\sigma_{23} + L_2\sigma_{13} + L_3\sigma_{12})^2
   = \frac{\lambda^{'}_{12}\lambda^{'}_{23}\lambda^{'}_{31}}{4}
    \left(\frac{r_1}{\lambda'_{23}} + \frac{r_2}{\lambda'_{13}} + \frac{r_3}{\lambda'_{12}}\right)^2\,,\label{VCB-nontriv}
\ee
which holds for {\em any} combination of the sign factors $\sigma_{ij}$. However, this expression is negative only in two cases: if exactly one among the three $\lambda'_{ij}$ is negative, and if all three $\lambda'_{ij}$ are negative. Thus, only in these two cases one needs to take this value into account when establishing the BFB conditions.

\subsection{The set of necessary and sufficient BFB conditions}
We are ready to formulate the set of necessary and sufficient BFB conditions for the $U(1) \times U(1)$ symmetric 3HDM, which we present as an algorithm.

{\bf Step 1.} 
Apply the copositivity conditions in \eqref{copositivity1} and \eqref{copositivity2} to 
the matrix $A = A_N$ extracted from $V_N$, eq.~\eqref{VN2}.

{\bf Step 2.} 
If at least one $\lambda'_{ij} < 0$, construct
three new matrices $A_{1,2,3} = A_N + \Delta_{1,2,3}$,
where $\Delta_{1,2,3}$ are extracted from the three expressions of the charge breaking potential $V_{CB}$
corresponding to the trivial solutions and listed in \eqref{VCB-triv}.
Apply the copositivity conditions in \eqref{copositivity1} and \eqref{copositivity2} to each of $A_{1,2,3}$.

\begin{figure}[H]
    \centering
    \includegraphics[width=.3\textwidth]{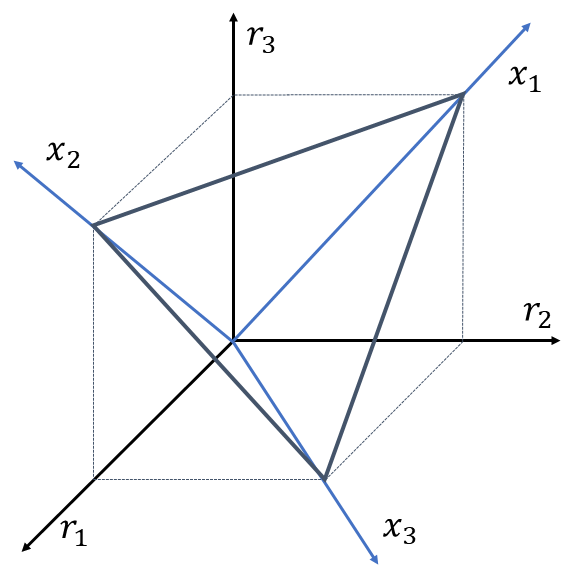}
    \caption{Location of the open tetrahedron defined 
    by $x_i \ge 0$ inside the first octant of $r_i$ space. Each axis $x_i$
    lies in the plane orthogonal to the axis $r_i$.}
    \label{figure}
\end{figure}

{\bf Step 3.}
If $\lambda^{'}_{12}\lambda^{'}_{23}\lambda^{'}_{31} < 0$, 
consider $V_N + V_{CB}^{\rm non-triv.}$, which is also a quadratic form of $r_i$,
and extract from it the new matrix $A_4 = A_N + \Delta_4$.
This matrix must be non-negative within the {\em open tetrahedron} in the $r_i$ space, illustrated by Fig.~\ref{figure}, 
that has the apex at the origin, lies inside the first octant, and is constrained by the triangle inequalities in \eqref{eq:prism}.
Non-negativity inside this tetrahedron can be achieved through the same copositivity technique.
%conditions applied to the 
Let us define new variables $x_{i}$, which are linearly related to $r_i$:
\begin{equation}
r_i = R_{ij} x_j\,,\qquad 
R=\begin{pmatrix} |\lambda'_{23}| & 0 & 0 \\ 0 & |\lambda'_{31}| & 0 \\ 0 & 0 & |\lambda'_{12}| \end{pmatrix} \begin{pmatrix} 0 & 1 & 1 \\ 1 & 0 & 1 \\ 1 & 1 & 0 \end{pmatrix}\,.
    \label{linear-map}
\end{equation}
Here, the first matrix links $r_i$ and $L_i$, while the second matrix alignes the axes $x_i$ with the directions of $L_i = 0$ (and therefore two other $L$'s being equal). The open tetrahedron defined by the inequalities in \eqref{eq:prism} corresponds to the first octant
in terms of the new variables: $x_i \ge 0$. Since the relation $r_i = R_{ij} x_j$ is linear, the quartic potential can be written as a quadratic form in the variables $x_i$ with the matrix $R^T A_4 R$. Therefore, to complete step 3 we need to check that the entries of this matrix satisfy the set of copositivity conditions in \eqref{copositivity1} and \eqref{copositivity2}.

These three steps represent the necessary and sufficient conditions for the potential of the $U(1) \times U(1)$ symmetric 3HDM to be bounded from below.

\section{Discussion}

\subsection{The necessity of step 3}

It may not be immediately obvious whether step 3 in the above strategy is indeed needed or whether it becomes redundant once steps 1 and 2 are passed. However, a quick check confirms that step 3 is indeed necessary.
Consider a very simple example: 
\begin{equation}
    \lambda_{11}=\lambda_{22}=\lambda_{33}=a > 0, \quad \lambda_{12}=\lambda_{13}=\lambda_{23}=0, \quad \lambda^{\prime}_{12}=\lambda^{\prime}_{13}=\lambda^{\prime}_{23}=- b,
\end{equation}
with a positive $b$.
Then, steps 1 and 2 result in the following conditions:
\begin{equation}
	a\ge 0\,, \quad a \ge b\,, \quad a \ge \sqrt{2}b\,. \label{simple-example}
\end{equation}
At step 3, one needs to add $V_{CB}^{\rm non-triv.}$ given by Eq.~\eqref{VCB-nontriv}. 
Instead of working out the copositivity conditions in full, let us just test the direction $r_1 = r_2 = r_3 = r$.
Evaluating the total quartic potential along this direction, one gets:
\begin{equation}
	V_4 = \frac{3}{2}r^2\left(a - \frac{3}{2}b\right)\,.
\end{equation} 
If one chooses $3/2 > a/b > \sqrt{2}$, the potential will pass steps 1 and 2 
but will be unbounded from below along this direction. Therefore, imposing step 3 is unavoidable.

\subsection{A pathological example}
It is important to appreciate that in certain cases the BFB conditions along the charge breaking directions in the Higgs space can be more constraining than along the neutral ones, and this is not related to the existence of a neutral minimum. Let us illustrate this point with a pathological example whose pathology would be easily missed if one focused only on the neutral directions.

Consider a $U(1) \times U(1)$ symmetric 3HDM model where the quadratic potential in eq.~\eqref{V2} contains only one negative coefficient: $m_{11}^2 = - |m_{11}^2|$. Then the minimum is at $\left<\phi_1^0\right> = v/\sqrt{2}$ with $v^2 = 2|m^2_{11}|/\lambda_{11}$, while $\left<\phi_{2,3}\right> = 0$. This is in fact the only phenomenologically viable choice, since it avoids spontaneous breaking of the $U(1)\times U(1)$ symmetry and does not lead to Goldstone bosons. It also leads to scalar dark matter candidates which is one of the main attraction points of symmetry protected multi-Higgs models. Expanding the potential around this point, one obtains masses of all physical scalars:
\begin{gather}
\begin{split}
    M^2_{h}=v^2\lambda_{11}, \quad & M^2_{H_2}=M^2_{A_2}=m^2_{22}+\frac{v^2}{2}\lambda_{12}, 
    \quad M^2_{H_3}=M^2_{A_3}=m^2_{33}+\frac{v^2}{2}\lambda_{13}, \nonumber\\
    & M^2_{H_2^\pm}=m^2_{22}+\frac{v^2}{2}(\lambda_{12}+\lambda^\prime_{12}), 
    \quad M^2_{H_3^\pm}=m^2_{33}+\frac{v^2}{2}(\lambda_{13}+\lambda^\prime_{13}).
\end{split}
\end{gather}
One can see that if $\lambda_{11}>0$ and if the quadratic parameters $m^2_{22},\,m^2_{33}$ are sufficiently large, 
then the masses squared are positive. Moreover, by choosing $\lambda_{11},\lambda_{22},\,\lambda_{33}>0$ and 
\begin{gather}
    \lambda_{12}>0, \quad \lambda_{13}>0, \quad \lambda_{23}>0,
\end{gather}
we can immediately guarantee that the potential is bounded from below in all neutral directions in the Higgs space.

However, these conditions are {\em not} sufficient to guarantee that the potential is bounded from below everywhere in the Higgs space. Take, for example, the following point in the parameter space:
\be
    \lambda_{11}=\lambda_{22}=\lambda_{33}=0.1, \quad \lambda_{12}=\lambda_{13}=\lambda_{23}=0.1, \quad \lambda^{\prime}_{12}=\lambda^{\prime}_{13}=\lambda^{\prime}_{23}=-0.6,
\ee
and explore how the quartic potential behaves along the ray\footnote{We do not claim that the potential is minimal along this direction. We simply show that there exists a direction along which the potential is unbounded from below.} $r_1=r_2=r_3\equiv r$, $\alpha_2=\alpha_3=\pi/2$. We find by direct calculation that $V_4=-0.75r^2$. It clearly indicates that the potential is unbounded from below despite the existence of a normally looking neutral minimum. Thus, in order to avoid such pathological models, one must always look for BFB conditions valid in the entire Higgs space, not only along the neutral directions. This remark if, of course, general and is not limited to the $U(1)\times U(1)$ symmetric 3HDM.

\subsection{Towards the BFB conditions in $\Z_2 \times \Z_2$ 3HDM}
The exploration of 3HDMs started 40 years ago with a model which combined natural flavor conservation (NFC) with various forms of $CP$ violation in the scalar sector \cite{weinberg1976gauge,branco1980spontaneous}. Indeed, NFC is imposed in this model by requiring that the potential be invariant under the global $\Z_2 \times \Z_2$ symmetry group generated, in a suitable basis, by independent sign flips of the doublets. The most general potential with this symmetry is given by the $U(1)\times U(1)$ symmetric potential $V_2 + V_N + V_{CB}$ in eqs.~\eqref{V2}, \eqref{VN1}, \eqref{VCB1} together with the additional phase-sensitive terms
\be
V_{\Z_2\times\Z_2} = \bar\lambda_{12} (\phi_1^\dagger\phi_2)^2 + 
\bar\lambda_{13} (\phi_1^\dagger\phi_3)^2 +
\bar\lambda_{23} (\phi_2^\dagger\phi_3)^2 + H.c., \label{VZ2Z2}
\ee
where $\bar\lambda_{ij}$ can be either real or complex.

When building models based on this potential, one must make sure that it is bounded from below. Despite the phenomenological interest raised by this model, for a long time there was no attempt to establish the exact necessary and sufficient BFB conditions for this model. Stability of the potential was only assumed, and the phenomenology was studied under this assumption. It was only in 2009 that this problem was addressed in \cite{grzadkowski2009natural}.  Results of that work, either for the generic $\Z_2 \times \Z_2$ quartic potential or under the assumption of ``dark democracy'', which sets some of the coefficients equal, were later used in several publications, e.g. \cite{Grzadkowski:2010au,Osland:2013sla,Moretti:2015tva,Cordero-Cid:2016krd}. Since the $U(1)\times U(1)$ symmetric 3HDM studied here is a particular case of the $\Z_2 \times \Z_2$ symmetric model, one would expect to be able to derive our BFB conditions from the results of Ref.~\cite{grzadkowski2009natural}. We found this not to be the case.

In order to clarify the situation, we introduce the notation used in Ref.~\cite{grzadkowski2009natural} and write the doublets as $\phi_i=||\phi_i||\hat{\phi_i}$, where the norms of the doublets are given by $||\phi_i|| = \sqrt{r_i}$ in our notation. The products of the unit doublets were written as
\begin{equation}
    \hat{\phi}^\dagger_2\hat{\phi}_1=\rho_1e^{i\varphi_1}, \quad \hat{\phi}^\dagger_3\hat{\phi}_1=\rho_2e^{i\varphi_2}, \quad \hat{\phi}^\dagger_3\hat{\phi}_2=\rho_3e^{i\varphi_3}\,,
\end{equation}
with $\rho_i \in [0,1]$ and $\varphi_i\in[0,2\pi)$. The whole quartic potential depends on them, but these 6 variables are themselves not independent. In general, three doublets of unit length $\hat{\phi_i} \in \mathbb{C}^2$ have 9 degrees of freedom. One can make use of the gauge symmetry to remove 4 of them, which leaves only 5 independent degrees of freedom. That is, these 6 variables must satisfy certain (in)equalities, which we derive by expressing these parameters with our parameterization in \cref{eq:Igor_parameterization}:
\begin{gather}
\rho_1=\cos{(\alpha_2)}, \quad \rho_2=\cos{(\alpha_3)}, \quad \varphi_1=-\beta_2, \quad \varphi_2=-(\beta_2+\gamma+\delta),
\end{gather}
 and 
\begin{equation}
    \hat{\phi}^\dagger_3\hat{\phi}_2=\left[\cos{(\alpha_2)}\cos{(\alpha_3)}e^{-i\delta}+\sin{(\alpha_2)}\sin{(\alpha_3)}\right]e^{-i\gamma}.
\end{equation}
 From the last relation we deduce that,
\begin{equation}\label{eq:rho_3}
    \rho^2_3=\rho^2_1\rho^2_2 + (1-\rho^2_1)(1-\rho^2_2) + 2\rho_1\rho_2\sqrt{(1-\rho^2_1)(1-\rho^2_2)}\cos{(\delta)},
\end{equation}
 which, for given $\rho_1$ and $\rho_2$, limits $\rho_3$ by
\begin{equation}\label{eq:rho_3 interval}
    \left|\rho_1\rho_2-\sqrt{(1-\rho^2_1)(1-\rho^2_2)}\right|\leq \rho_3 \leq \rho_1\rho_2+\sqrt{(1-\rho^2_1)(1-\rho^2_2)}.
\end{equation}
Thus, when minimizing the potential in the $\rho_i$ space, one must take into account that the space available is not the full cube $0 \le \rho_i \le 1$, but a part of it bounded by the inequalities in \eqref{eq:rho_3 interval}.

Moreover, the 3 phases $\varphi_i$ are not independent. Let us construct the rephasing-invariant quantity
\begin{equation}
    \left(\hat{\phi}^\dagger_2\hat{\phi}_1\right) \left(\hat{\phi}^\dagger_1\hat{\phi}_3\right) \left(\hat{\phi}^\dagger_3\hat{\phi}_2\right)=\rho_1\rho_2\rho_3e^{i\Sigma\varphi}=\rho_1\rho_2\left[\cos{(\alpha_2)}\cos{(\alpha_3)}+\sin{(\alpha_2)}\sin{(\alpha_3)}e^{i\delta}\right]\,.
\end{equation}
Here $\Sigma\varphi \equiv\varphi_1-\varphi_2+\varphi_3$ can be expressed as 
\begin{equation}
    \tan{\left(\Sigma\varphi\right)}=\frac{\sin{(\alpha_2)}\sin{(\alpha_3)\sin{(\delta)}}}{\cos{(\alpha_2)}\cos{(\alpha_3)}+\sin{(\alpha_2)}\sin{(\alpha_3)\cos{(\delta)}}}\,.
\end{equation}
Switching back to the $\rho_i$ notation and using \cref{eq:rho_3}, we find after some algebra
\begin{equation}\label{eq:sum_varphi}
    \cos^2\left(\Sigma\varphi\right)=\frac{\left(\rho^2_1+\rho^2_2+\rho^2_3-1\right)^2}{4\rho^2_1\rho^2_2\rho^2_3}.
\end{equation}
From this expression it is clear that, for given $\rho_i$, the quantity $\Sigma\varphi=\varphi_1-\varphi_2+\varphi_3$ is fixed, up to discrete ambiguities. 

By treating all 6 variables $\rho_{1,\,2,\,3}$ and $\varphi_{1,\,2,\,3}$ as independent and minimizing the potential with respect to them individually, as was done in Ref.~\cite{grzadkowski2009natural}, one would arrive at a value of the potential which would be {\em lower} than what actually is possible to achieve within the space available. Thus, the conditions obtained are, at most, sufficient but not necessary. It is possible to construct examples of this model which violate the previously found conditions but whose potentials are, nevertheless, bounded from below. It would be interesting to see if these parts of the allowed, but neglected, parameter space correspond to any characteristic phenomenology.

\section{Conclusions}
Exploration of viable parameter space regions in models with extended Higgs sectors can be challenging due to sophisticated scalar potentials. In particular, requiring that the potential be bounded from below (BFB) places constraints on its parameters, but establishing the exact necessary and sufficient BFB conditions can be notoriously difficult. In this paper we found these conditions in the rephasing invariant three-Higgs-doublet model, which can be used to construct viable models with degenerate scalar dark matter candidates.

When deriving these conditions, we found that it is extremely important to check not only neutral but also charge-breaking directions in the Higgs space. To highlight this point, we showed an example of this model which possesses a good-looking minimum, with positive masses squared for all Higgses, and whose potential is bounded from below in all neutral directions of the Higgs space. Yet the example is pathological because the potential is not bounded from below in charge-breaking directions. 
One may argue that if this metastable vacuum is sufficiently long-lived, one can still develop phenomenology around it.
We do not enter this discussion here; we only give a direct proof that having a neutral minimum 
is not an excuse to sidestep stability checks in the entire Higgs space.

It would be even more interesting to find the exact BFB conditions for the $\Z_2 \times \Z_2$ symmetric case,
the model with which 3HDMs made their debut in late 1970s \cite{weinberg1976gauge,branco1980spontaneous}. 
This model includes, in addition to our potential, extra terms. 
Our results can be seen as a set of necessary conditions for this case, but they are not sufficient.
We have not yet solved this problem, but we notice that the approach taken in Ref.~\cite{grzadkowski2009natural} does not lead to the necessary and sufficient BFB conditions for the $\Z_2 \times \Z_2$ symmetric 3HDM. 
This problem remains to be solved.

\section*{Acknowledgements}
We thank Jo\~{a}o~P.\ Silva and Andreas Trautner for many useful discussions and comments on the paper. We also thank Bohdan Grzadkowski, Per Osland, and Odd Magne Ogreid for correspondence.
We acknowledge funding from the Portuguese \textit{Fun\-da\-\c{c}\~{a}o para a Ci\^{e}ncia e a Tecnologia} (FCT) through the FCT Investigator contracts IF/00989/2014/CP1214/CT0004 and IF/00816/2015 and through the projects UID/FIS/00777/2013, UID/FIS/00777/2019, CERN/FIS-PAR/0004/2017, and PTDC/FIS-PAR/29436/2017, which are partially funded through POCI, COMPETE, QREN, and the EU. We also acknowledge the support from National Science Center, Poland, via the project Harmonia (UMO-2015/18/M/ST2/00518).

\bibliography{BibTexFile}

\end{document}